# Large Scale Enrichment and Statistical Cyber Characterization of Network Traffic

Ivan Kawaminami[1], Arminda Estrada[1], Youssef Elsakkary[1], Hayden Jananthan[2], Aydın Buluç[3], Tim Davis[4], Daniel Grant[5], Michael Jones[2], Chad Meiners[2], Andrew Morris[5], Sandeep Pisharody[2], Jeremy Kepner[2]
[1]University of Arizona, [2]MIT, [3]LBNL, [4]Texas A&M, [5]GreyNoise,

*Abstract—* **Modern network sensors continuously produce enormous quantities of raw data that are beyond the capacity of human analysts. Cross-correlation of network sensors increases this challenge by enriching every network event with additional metadata. These large volumes of enriched network data present opportunities to statistically characterize network traffic and quickly answer a key question: "What are the primary cyber characteristics of my network data?" The Python GraphBLAS and PyD4M analysis frameworks enable anonymized statistical analysis to be performed quickly and efficiently on very large network data sets. This approach is tested using billions of anonymized network data samples from the largest Internet observatory (CAIDA Telescope) and tens of millions of anonymized records from the largest commercially available background enrichment capability (GreyNoise). The analysis confirms that most of the enriched variables follow expected heavy-tail distributions and that a large fraction of the network traffic is due to a small number of cyber activities. This information can simplify the cyber analysts' task by enabling prioritization of cyber activities based on statistical prevalence.**

## I. Introduction

"What are the primary cyber characteristics of my network data?" is a critical question for anyone protecting a network. Understanding network data characteristics statistically is a powerful tool for prioritizing resources for the most significant cyber activities. Cyber characterization often requires the analysis and correlation of significant volume of network traffic often containing billions of (anonymized) records.

Internet sensors enable the collection of the raw data used to characterize network traffic to help organizations protect cyber assets [1]. This work exhibits large-scale statistical cyber characterization using datasets from the Center for Applied Internet Data Analysis (CAIDA) Telescope and the GreyNoise honeyfarm that is cross-correlated using GraphBLAS (GraphBLAS.org) and Python Dynamic Distributed Dimensional Data Model (d4m.mit.edu) libraries [2].

The CAIDA Telescope passively monitors Internet packets into an internet darkspace that is a globally routed /8 network that is almost entirely adversarial traffic [3]. Honeyfarms are a mechanism for actively observing and interacting with internet traffic to create richer data sets [4]. The GreyNoise honeyfarm consists of hundreds of servers passively collecting packets from IPs scanning the internet. GreyNoise servers have the additional capability to converse with the IP sources to identify behavior, methods, and intent [5,6]. A significant difference between these two datasets is that CAIDA collects significantly more data over the same period than GreyNoise does – approximately thirty minutes of CAIDA data collection is equivalent to thirty days of GreyNoise data collection. Table I depicts the collected data used for this research. The data from Table I has two start times, GreyNoise and CAIDA. Throughout this paper, the CAIDA start time is used to specify the data collected dates, as seen in Tables II-VI.

Table I. Data collection start time, collection duration, and the number of unique sources from the GreyNoise and CAIDA datasets. GreyNoise data was collected for a month, while CAIDA data was approximately every six weeks on Wednesdays, either at noon or midnight. Constant packet and variable time samples simplify the statistical analysis of the heavy-tail distributions commonly found in network traffic quantities [7].

| GreyNoise Start Time | GreyNoise Duration | GreyNoise Sources | CAIDA Start Time | CAIDA Duration | CAIDA Packets | CAIDA Sources |
|---|---|---|---|---|---|---|
| 2020-06-01 | 30 days | 1,111,458 | 20200617-120000 | 1594 sec | $2^{30}$ | 670,304 |
| 2020-07-01 | 31 days | 1,438,698 | 20200729-000000 | 1312 sec | $2^{30}$ | 541,300 |
| 2020-09-01 | 30 days | 1,245,194 | 20200916-120000 | 997 sec | $2^{30}$ | 723,991 |
| 2020-10-01 | 31 days | 1,997,782 | 20201028-000000 | 1068 sec | $2^{30}$ | 796,327 |

The datasets contain network traffic found by collecting data packets from thousands of Internet Protocol (IP) addresses that scan the internet daily. Previous work [7] examined the temporal correlation between CAIDA sources seen by GreyNoise over time, finding that among CAIDA sources that have sent $d < N_V^{1/2}$ source packets. A modified Cauchy distribution describes the fraction found in GreyNoise sources

$$p(d,t) \propto \frac{\log_2(d)}{\log_2(N_V^{1/2})} \frac{\beta}{\beta + |t - t_0|^\alpha}$$

where $t$ is the GreyNoise measurement time, $t_0$ is the CAIDA measurement time, $N_V$ is the number of CAIDA packets, and $\alpha$ and $\beta$ are empirically determined model parameters. In general, 70% of the highest frequency ($d > N_V^{1/2}$) sources in CAIDA are also seen in the same GreyNoise time window.

Building on this prior work, this paper explores in detail the cyber characteristics of the correlated sources that are seen in both data sets. Understanding these characteristics can aid in defending against cyberattacks. The CAIDA telescope darkspace data is rich starting point for this characterization as it represents almost entirely adversarial traffic [8-10]. Previous works made in collaboration with different data collection centers, such as the Widely Integrated Distributed Environment (WIDE) project and CAIDA, have used such characterizations to understand the global state of Internet traffic and test prototypes to stop internet worms [11,12].

In related work, network characterization can leverage machine learning for anomaly detection, intrusion detection, and behavior analysis[13]. Some machine learning classifiers identify malicious activity prediction with 96% accuracy for intrusion detection even in zero-day exploited vulnerabilities in internet traffic [14]. Likewise, Internet traffic data can be categorized into audio-stream, browsing, P2P, file-transfer, video-stream, and VOIP types based on representative classification features, such as flow duration, time intervals among packets, packet statistics, number of packets, inbound packets, outbound packets, …, with different analysis applied to each [15-19]. These approaches include convolutional neural networks and bidirectional long short-term memory neural networks to capture local features from the packet's content. Such approaches have resulted in multiclass accuracy of 92% even for encrypted traffic types [20].

CAIDA has been further used to understand internet traffic, in [21-24] addressing the problem of locating the exact Autonomous System Number (ASN), which controls internet routing. In this case, it discriminates traffic by analyzing all possible AS levels and paths by combining measurements from CAIDA; besides analyzing big data traffic using machine learning, it is possible to create systems for intrusion detection. Furthermore, in [25], traffic obtained from CAIDA was used to observe virus and worm behavior over time by applying the approaches mentioned previously.

Prior work has focused on the specific classification of network traffic data, but due to computational limitations, broad statistical characterization on a large scale has been a challenge. This paper demonstrates the feasibility of such characterization on corpora with billions of packets and 100,000s of sources (see Table I). Furthermore, anonymized Internet Protocol (IP) addresses are used to demonstrate that such characterizations can be done with privacy considerations. Working with anonymized IP addresses becomes a powerful aid in data sharing. Anonymizing the IP addresses enables the sharing of the data while keeping the user's information private. The Python GraphBLAS and PyD4M analysis frameworks are key tools to efficiently and effectively perform this analysis on large amounts of anonymized data.

This research addresses the characterization of the cross-correlated CAIDA/GreyNoise data in the following order: Section II presents a dimensional data analysis to determine the variables of interest. Section III develops an "exemplar" record from the most common variables. Section IV computes the probability distributions of selected variables and models them with a 2-parameter Zipf-Mandelbrot distribution. Finally, Section V presents the conclusions of the analysis and future work.

## II. Dimensional Data Analysis

Dimensional analysis is used to observe relationships between quantities based on their dimensions in complex systems. Such analysis can be used to understand the information content of the databases and find inconsistencies, data patterns, and formatting. Relationships do not change when the units of measurement are altered [26]. This concept has been extended to "big data" in the form of dimensional data analysis which looks at size (or dimension) of different values in the data [27]. In this paper, dimensional data analysis of the CAIDA-GreyNoise data helps identify those variables that are amenable to further exploration. Given the amount and type of data being handled, the Python library D4M is used, utilizing the concepts of linear algebra and signal processing to databases through associative arrays while providing a schema capable of representing most data and providing a low barrier to entry through a simple API.

The CAIDA-GreyNoise data is constructed by building GraphBLAS hypersparse traffic matrices for each $2^{30}$ packet CAIDA sample set listed in Table I. The logarithmically binned number of **source packets**, $\log_2(d)$, from each CAIDA source are computed and the source IPs are anonymized using the CryptoPAN protocol. Likewise, the source IPs from the GreyNoise are similarly anonymized allowing the two data sets two be cross-correlated.

The combined CAIDA-GreyNoise variables associated with each anonymized source IP are then separated into three tables based on their dimensional data analysis types: one containing variables that did not need further analysis and two tables with variables that are amenable to further analysis. The latter two tables correspond to how many unique values are associated with those variables, either high (~10,000 or more, Table IV) or low (~10, Table V). The three tables have the same general format with column headings:

**Date:** The start date and time of the CAIDA collection, approximately every six weeks on Wednesday, either at noon (120000) or midnight (000000).

**Variable:** The variable of interest for analysis, collected by CAIDA and GreyNoise.

**Nrow:** Number of source IPs associated with this variable.

**Ncol:** Number of unique values for specific variables.

**Nnz:** Number of nonzero values.

**Maxval:** The value appearing most frequently.

**Maxcount:** The number of times **maxval** appears.

**Maxfrac:** The fraction of the number of times the maximum value appears over each variable's total number of values (i.e., Maxfrac = Maxcount/Ncol). *Maxfrac* is expressed with italics to indicate it can be determined directly from Maxcount and Ncol.

From the data provided by CAIDA and GreyNoise, the variables chosen for analysis include:

**Actor:** A participant sending the packets over the network.

**ASN:** Autonomous System Number is the number that identifies each autonomous system. It is a collection of connected internet protocol routing prefixes controlled by network operators.

**Classification:** GreyNoise classification of the source IP as unknown, benign, and malicious. A device can be classified as "malicious" if it is flagged as having malicious intent, "benign" if it is without malicious intent, and "unknown" if the purpose of the transmitted packets are unknown.

**CVE:** Common Vulnerabilities and Exposures (CVE) lists publicly known vulnerabilities in devices with related information. Records classified as "benign" or "unknown" do not have a CVE value, while each "malicious" record may have zero or more CVEs associated with it.

**Last seen timestamp** (last seen): The timestamp when GreyNoise last observed the source IP during the one-month observation window.

**OS:** The Operating system (OS) associated with the source IP.

**Protocol port:** network ports associated with communications from the source IP. The port list was truncated at 5 to avoid excessive numbers of ports in the analysis.

**Source packet** (srcPacket): The logarithmically binned $\log_2(d)$ number of packets from each CAIDA source.

**Source packet No Grey** (caida srcPacket): Same as **source packet** but for those CAIDA source IPs that are *not* found in the GreyNoise data.

**CAIDA Grey Meta** (caidaGreyMeta)/**CAIDA No Grey** (caidaNoGrey): The total values in the matrix for each file, where caidaGrayMeta is the data correlated between GreyNoise and CAIDA and caidaNoGrey is solely obtained from CAIDA.

**Spoofable:** This binary variable identifies whether the source IP has failed to complete a full connection or not.

*A. Source Variables*

Table II is the source table containing more detailed information on the amount of data obtained. The date column shows four dates beginning with the month's year, month, and day number. If a date ends in "120000" the data was collected by CAIDA at noon; however, if the date ends with "000000", the data was collected at midnight.

Table II. Source table for the data collected on four dates; divided into the year, month, and day of the month. Variables caidaGreyMeta corresponds to source IPs appearing in both GreyNoise and CAIDA datasets, whereas caidaNoGrey corresponds to source IPs appearing in only CAIDA's dataset.

| date | variable | nrow | ncol | nnz | max count | maxfrac |
|---|---|---|---|---|---|---|
| 20200617-120000 | caidaNoGrey | 545,980 | 1 | 545,980 | 545,980 | 1 |
| 20200729-000000 | caidaNoGrey | 394,875 | 1 | 394,875 | 394,875 | 1 |
| 20200916-120000 | caidaNoGrey | 589,267 | 1 | 589,267 | 589,267 | 1 |
| 20201028-000000 | caidaNoGrey | 603,273 | 1 | 603,273 | 603,273 | 1 |
| 20200617-120000 | caidaGreyMeta | 124,300 | 9 | 942,112 | 124,300 | 0.131 |
| 20200729-000000 | caidaGreyMeta | 146,402 | 9 | 1,118,893 | 146,402 | 0.130 |
| 20200916-120000 | caidaGreyMeta | 134,699 | 9 | 1,034,353 | 134,699 | 0.130 |
| 20201028-000000 | caidaGreyMeta | 193,029 | 9 | 1,470,649 | 193,029 | 0.131 |

**caidaNoGrey/caidaGreyMeta:** Table II contains the metadata information for CAIDA-GreyNoise cross-correlated records and metadata for CAIDA records with no associated GreyNoise metadata. For caidaGreyMeta, there are nine distinct values representing the variables being analyzed: actor, spoofable, ASN, last seen timestamp, protocol port, os, srcPacket, classification, and CVE. On the other hand, caidaNoGrey shows the one variable caida srcPacket.

*B. Irrelevant Variables*

Table III depicts the variables with a limited number of distinct values that are unlikely to yield additional information.

**Spoofable:** Table III shows a many-to-one correlation between the spoofable packets and the Boolean value "1". Between 30% - 40% of the source IPs are spoofable though being flagged as spoofable does not necessarily imply that the traffic was spoofed, only that the inverse could not be established. There does not appear to be any major correlation between being flagged as spoofable and the remaining variables. This variable is then irrelevant for further analysis.

**Actor:** Table III shows that there are between 124,300 and 193,029 records found for the actors with a many-to-one correlation; in other words, for each packet, there is a single actor (though this actor may be "unknown"). There are between 22 to 40 unique actors, with the most common actor being "unknown", showing up about 99% of the time. Because so few records have a 'known' actor, this variable is not relevant.

Table III. Irrelevant variables that are unlikely to benefit from further analysis. Variables include actor and spoofable, with the maximum value of actor being "unknown" for any date, while spoofable is a simple binary value.

| date | variable | nrow | ncol | nnz | maxval | max count | maxfrac |
|---|---|---|---|---|---|---|---|
| 20200617-120000 | spoofable | 36,317 | 1 | 36,317 | 1 | 36,317 | 1 |
| 20200729-000000 | spoofable | 43,838 | 1 | 43,838 | 1 | 43,838 | 1 |
| 20200916-120000 | spoofable | 56,189 | 1 | 56,189 | 1 | 56,189 | 1 |
| 20201028-000000 | spoofable | 62,172 | 1 | 62,172 | 1 | 62,172 | 1 |
| 20200617-120000 | actor | 124,300 | 22 | 124,300 | unknown | 123,843 | 0.996 |
| 20200729-000000 | actor | 146,402 | 26 | 146,402 | unknown | 145,812 | 0.996 |
| 20200916-120000 | actor | 134,699 | 30 | 134,699 | unknown | 133,036 | 0.987 |
| 20201028-000000 | actor | 193,029 | 29 | 193,029 | unknown | 192,091 | 0.995 |

*C. Relevant Variables*

Table IV and Table V are separated by variables with statistical similarities, the size of the number of columns, and, particularly, the number of unique values in the records. Table IV contains the variables with many unique values in the thousands, while Table V contains several unique values in the two digits. Variables in both tables are subsequently analyzed using probability distributions and histograms.

Table IV. Variables related to many (~10,000 and above) unique values, the number of which is represented by ncol. Variables include ASN, last seen timestamp, and port protocol.

| date | variable | nrow | ncol | nnz | maxval | max count | maxfrac |
|---|---|---|---|---|---|---|---|
| 20200617-120000 | protocol port | 124,300 | 4,021 | 203,081 | TCP/445 | 44,607 | 0.219 |
| 20200729-000000 | protocol port | 146,402 | 4,353 | 219,650 | TCP/445 | 66,155 | 0.301 |
| 20200916-120000 | protocol port | 134,699 | 6,075 | 210,315 | TCP/23 | 46,874 | 0.223 |
| 20201028-000000 | protocol port | 193,029 | 5,541 | 304,387 | TCP/445 | 78,971 | 0.259 |
| 20200617-120000 | ASN | 124,300 | 10,304 | 124,300 | AS4134 | 6,096 | 0.049 |
| 20200729-000000 | ASN | 146,402 | 9,825 | 146,402 | AS3462 | 8,650 | 0.059 |
| 20200916-120000 | ASN | 134,699 | 10,059 | 134,699 | AS17488 | 130,24 | 0.097 |
| 20201028-000000 | ASN | 193,029 | 10,764 | 193,029 | AS4837 | 118,30 | 0.061 |
| 20200617-120000 | last seen | 124,300 | 115,900 | 124,300 | 6/23/20 23:20 | 18 | 0.000143 |
| 20200729-000000 | last seen | 146,402 | 130,474 | 146,402 | 8/1/20 0:09 | 16 | 0.000109 |
| 20200916-120000 | last seen | 134,699 | 127,504 | 134,699 | 10/1/20 0:11 | 10 | 0.000074 |
| 20201028-000000 | last seen | 193,029 | 157,549 | 193,029 | 10/31/20 23:12 | 40 | 0.000207 |

**Protocol port:** At times, many protocol ports were used for single-packet transmission by a Source IP. The number of protocol ports per record was limited to five for the feasibility of working with the data. The number of unique protocol ports was between 40,000 and 61,000. TCP/445 (SMB) was the most frequent on three of the collection dates, and TCP/23 (telnet) on 20200916-120000. The most popular protocol port appeared between 21% and 30% of the time.

**ASN:** Each record has one ASN associated with it. There are thousands of distinct ASNs, and there is a different maximum value for each date. The times those maximum values are counted are between 6,096 and 13,024 times, corresponding to between 4.9% and 6.1% of the total of ASNs.

**Last seen timestamp (last seen):** For each given record, there is exactly one last seen timestamp. There are between 115,900 and 157,549 unique last seen timestamps. The maximum values, as expected, are related to the date and time the data was collected. The range count of maximum timestamps is between 10 to 40 or less than 0.02% of the total number of timestamps recorded.

Table V contains variables that are further analyzed. The similarities between these variables are defined by the number of unique values (ncol) being fairly low – between 3 to 36 unique values each.

Table V. Variables exhibiting relatively few unique values (~10), including classification, CVE, os, srcPacket, and srcPacketnoGrey.

| date | variable | nrow | ncol | nnz | maxval | max count | Max frac |
|---|---|---|---|---|---|---|---|
| 20200617-120000 | CVE | 35,695 | 36 | 37,150 | CVE-2017-0144 | 33,660 | 0.906 |
| 20200729-000000 | CVE | 50,241 | 33 | 51,546 | CVE-2017-0144 | 48,265 | 0.936 |
| 20200916-120000 | CVE | 35,271 | 32 | 36,500 | CVE-2017-0144 | 32,205 | 0.882 |
| 20201028-000000 | CVE | 57,274 | 34 | 59,252 | CVE-2017-0144 | 52,880 | 0.892 |
| 20200617-120000 | classification | 124,300 | 3 | 124,300 | malicious | 72,952 | 0.587 |
| 20200729-000000 | classification | 146,402 | 3 | 146,402 | malicious | 86,058 | 0.588 |
| 20200916-120000 | classification | 134,699 | 3 | 134,699 | malicious | 86,141 | 0.639 |
| 20201028-000000 | classification | 193,029 | 3 | 193,029 | malicious | 109,849 | 0.569 |
| 20200617-120000 | os | 124,300 | 17 | 124,300 | Windows 7/8 | 45,107 | 0.363 |
| 20200729-000000 | os | 146,402 | 20 | 146,402 | Windows 7/8 | 68,856 | 0.470 |
| 20200916-120000 | os | 134,699 | 31 | 134,699 | Windows 7/8 | 45,502 | 0.338 |
| 20201028-000000 | os | 193,029 | 20 | 193,029 | Windows 7/8 | 83,013 | 0.430 |
| 20200617-120000 | caida srcPacket | 545,980 | 23 | 545,980 | 32 | 104,466 | 0.191 |
| 20200729-000000 | caida srcPcaket | 394,875 | 24 | 394,875 | 32 | 67,020 | 0.169 |
| 20200916-120000 | caida srcPacket | 589,267 | 22 | 589,267 | 1 | 113,371 | 0.1925 |
| 20201028-000000 | caida srcPacket | 603,273 | 24 | 603,273 | 1 | 105,623 | 0.175 |
| 20200617-120000 | srcPacket | 124,300 | 25 | 124,300 | 128 | 20,570 | 0.165 |
| 20200729-000000 | srcPacket | 146,402 | 24 | 146,402 | 64 | 19,750 | 0.135 |
| 20200916-120000 | srcPacket | 134,699 | 26 | 134,699 | 64 | 21,907 | 0.166 |
| 20201028-000000 | srcPacket | 193,029 | 25 | 193,029 | 64 | 31,783 | 0.165 |

**CVE:** Between 35,271 and 57,274 records have a reported CVE; that is between 26% and 30% of the malicious traffic represented, which is less than half of the records. Many CVEs may appear in a single record. The most popular CVE across all dates is CVE-2017-0144. It is a well-known vulnerability called Eternal Blue related to the famous WannaCry attack that became a global problem in 2017 [28]. Between 88% and 93% of the time a CVE was reported.

**Classification:** There are three unique values – "malicious", "benign", and "unknown" – for which a "malicious" value represents between 59.6% and 63.95% (72,952 to 109,849 of the records).

**OS:** The number of records with a reported operating system is between 124,300 and 193,029. Every record has a recorded operating system, though this includes "unknown", which indicates GreyNoise failed to determine the Source IPs OS. There is a many-to-one relationship between source IPs and OS, with several unique values ranging from 17 to 31 (including "unknown"); see Table IV. The most common OS is Windows 7/8, with a maximum count ranging from 45,107 to 83,013, representing between 33% and 47% of those reported.

**Source packet No Grey (caida srcpacket):** Records with reported source packets range from 39,487 to 60,327. Table IV depicts a many-to-one correlation between source IPs and the (binned) number of source packets found in the CAIDA records. The most common source packet bin size is 1 for two dates and 32-63 for the other two dates. The maximum number of times a specific source packet size range appears is from 67,020 to 113,371 times or 16.98% to 19.24% of the time.

**Source packet (src packet):** Records with reported source packets are between 124,300 and 193,029. Unique sources binned in sizes $2^n$ range from 23 and 26. Table V exhibits a many-to-one correlation between source IPs and source packet bins; for every entry, there is exactly one source packet range found. The most common source packet size is between 64 and 127 kilobytes and, on one occasion, between 128 and 256 kilobytes. Between 13% to 16.5% of the packets belong to this size range or between 19,750 to 31,783 records.

## III. EXEMPLAR RECORDS

Based on the dimensional data analysis, it is possible to provide an initial answer to the question of "What are the primary cyber characteristics of my network data?" in the form of an exemplar record based on the data's most common values.

Table VI. Exemplar records reflecting the most frequent value of each variable.

| variable | maxval | maxcount | maxfrac | date |
|---|---|---|---|---|
| actor | unknown | 123843 | 0.996 | 20200617-120000 |
| | | 145812 | 0.996 | 20200729-000000 |
| | | 133036 | 0.988 | 20200916-120000 |
| | | 192091 | 0.995 | 20201028-000000 |
| CVE | CVE-2017-0144 | 33660 | 0.906 | 20200617-120000 |
| | | 48265 | 0.936 | 20200729-000000 |
| | | 32205 | 0.882 | 20200916-120000 |
| | | 52880 | 0.892 | 20201028-000000 |
| classification | malicious | 72952 | 0.587 | 20200617-120000 |
| | | 86058 | 0.588 | 20200729-000000 |
| | | 86141 | 0.640 | 20200916-120000 |
| | | 109849 | 0.569 | 20201028-000000 |
| OS | Windows 7/8 | 45107 | 0.363 | 20200617-120000 |
| | | 68856 | 0.470 | 20200729-000000 |
| | | 45502 | 0.338 | 20200916-120000 |
| | | 83013 | 0.430 | 20201028-000000 |
| protocol port | TCP/445 | 44607 | 0.220 | 20200617-120000 |
| | | 66155 | 0.301 | 20200729-000000 |
| | TCP/23 | 46874 | 0.223 | 20200916-120000 |
| | TCP/445 | 78971 | 0.259 | 20201028-000000 |
| source packet noGrey | 32 | 104466 | 0.191 | 20200617-120000 |
| | | 67020 | 0.170 | 20200729-000000 |
| | 1 | 113371 | 0.192 | 20200916-120000 |
| | | 105623 | 0.175 | 20201028-000000 |
| source packet | 128 | 20570 | 0.165 | 20200617-120000 |
| | 64 | 19750 | 0.135 | 20200729-000000 |
| | | 21907 | 0.163 | 20200916-120000 |
| | | 31783 | 0.165 | 20201028-000000 |
| ASN | AS4134 | 6096 | 0.049 | 20200617-120000 |
| | AS3462 | 8650 | 0.059 | 20200729-000000 |
| | AS17488 | 13024 | 0.097 | 20200916-120000 |
| | AS4837 | 11830 | 0.061 | 20201028-000000 |

Table VI the highly frequent values among the variables. An exemplar record based on maxval and maxcount would be 99% of the time, an unknown actor running the Windows 7/8 operating system sending between 64 and 127 packets with malicious intent over TCP/455 with a nonspecific ASN and targets CVE-2017-0144. Such an exemplar record provides a network operators' with highly actionable information about the most common adversarial activities on their network.

## IV. PROBABILITY DISTRIBUTIONS

While an exemplar record provides an indication of the most common activities, probability distributions shed insight into the broader behavior of the network. Furthermore, many of the variables are consistent with a Zipf-Mandelbrot distribution which allows the characterization of the data with just 2 parameters [29]. The Zipf-Mandelbrot model is denoted by

$$p(d; \alpha, \delta) \propto \frac{1}{(d+\delta)^\alpha}$$

Where d is the count of the particular value of the variable and $\delta$ and $\alpha$ are parameters found using the procedures of [7]. In Figures 1-4, the computed modified Zipf-Mandelbrot model is plotted, noting that the model $p(d; \alpha, \delta)$ is logarithmically binned before plotting. Figure 1 shows that most source IPs of the network sent around $10^2$ packets while very few source IPs sent a large number of packets.

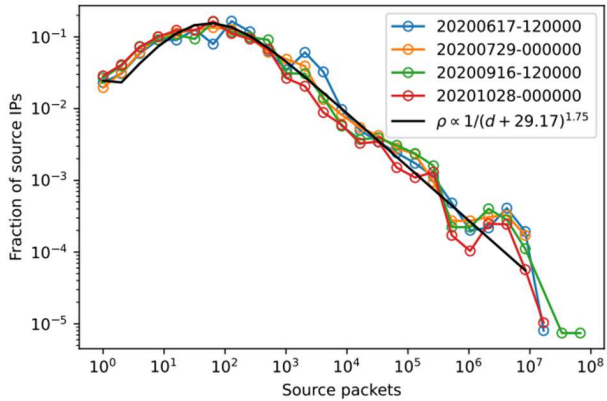

Figure 1. Probability distribution of the fraction of CAIDA source IPs (without GreyNoise matches) versus the number of source packets from the source IP. Most source IPs transmitted between $10^1$ and $10^3$ packets. The data are consistent with a Zipf-Mandelbrot distributions with α=1.75 and δ=29.17.

Figure 2 has similar behavior to that of Figure 1, in line with the fact that CAIDA and GreyNoise are each observing the similar internet traffic.

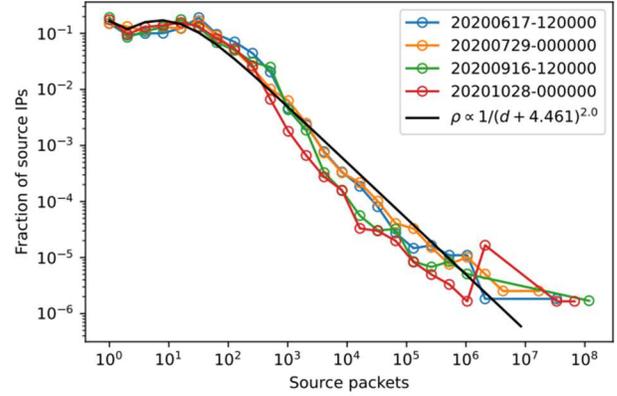

Figure 2. Probability distribution of the fraction of CAIDA source IPs (with GreyNoise matches) versus the number of source packets from the source IP. Most source IPs transmitted less than $10^2$ packets. The data are consistent with a Zipf-Mandelbrot distributions with α=2.0 and δ=4.461. The distribution shows a high number of source IPs with a low number of source packets and a low number of source IPs with a high number of source packets.

The histogram in Figure 3 shows that low-frequency protocol ports represent a higher fraction of the data versus high-frequency ports, suggesting that uncommon protocol ports are favored instead of high-frequency protocol ports by illegitimate traffic.

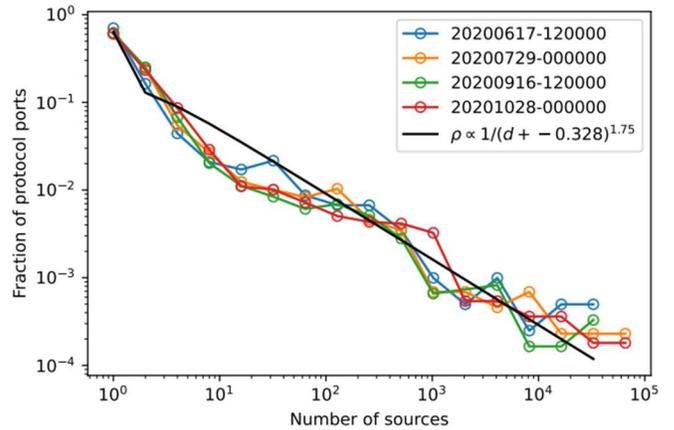

Figure 3. Probability distribution of the fraction of protocol ports versus the number of sources with that protocol port. The data are consistent with a Zipf-Mandelbrot distributions with α=1.75 and δ=-0.328. Probability distribution of the number of source IPs versus the fraction of protocol ports utilized by that number of sources. The distribution shows a high number protocol ports used by only a few sources and few protocol ports used by many sources.

The same behavior is seen with ASNs, as in Figure 4, suggesting that uncommon ASNs are favored by illegitimate traffic. Figures 2-4 suggest these variables are relatively consistent over time and consistent with the Zipf-Mandelbrot distributions. These data allow a network operator to prioritize sources based on their impact. Likewise, comparison of new data with the expected distribution can be used to detect changes in behavior which may suggest additional measures.

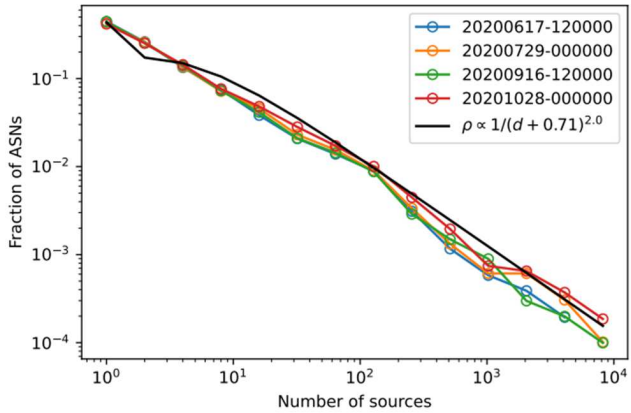

Figure 4. Probability distribution of the fraction of ASNs versus the number of sources with that ASN. The data are consistent with a Zipf-Mandelbrot distributions with α=2.0 and δ=0.72. The distribution shows many ASNs were used by only a few sources and few ASNs were used by many sources.

In Figure 5, the correlation between the hour component of the last seen timestamp and the fraction of source IPs last seen at that hour of the day is shown. Lines with 120000 in their label mean the collection started at noon, and those labeled with 000000 were collected at midnight. Data collected at the same time of the day show a close correlation.

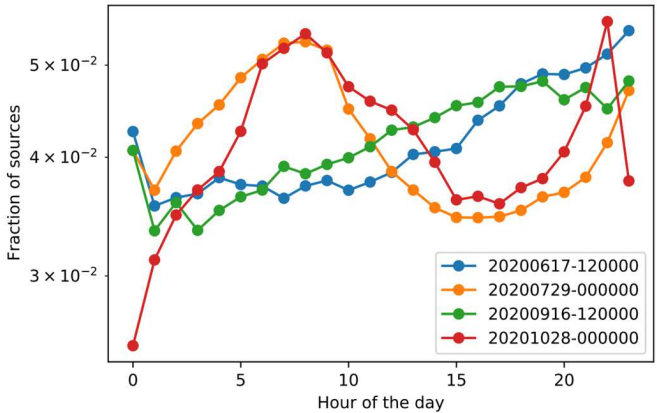

Figure 5. Probability distribution of the hour of the day versus the fraction of source IPs whose last seen timestamp was at that time of day. For the dates ending in 120000, the CAIDA data was collected at noon; for dates ending in 000000, the collection was at midnight. The distributions of last seen timestamps are consistent across those with the same collection times of 120000 or 000000.

The classification distribution in Figure 6 shows that among the three unique values— "benign", "malicious", and "unknown" –there are almost no benign packets. Rather, the majority are malicious, and the rest are unknown. The relative frequency is consistent across the four-collection period.

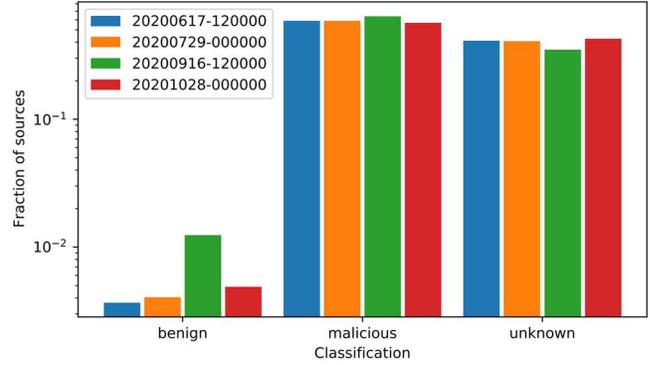

Figure 6. Packet type frequency for CAIDA and GreyNoise. There were three types of classification, with the majority classification for each date being malicious.

Figure 7 shows the distribution of the OSes that is relatively consistent across the four collection periods. While Windows 7/8 was the most common OS from the source IPs, the second most common OS was unknown, followed closely by several versions of Linux. Figure 8 shows a similar trend in time. Both Figure 7 and Figure 8 corroborate the exemplar record shown in Section III, with the most common operating system being Windows 7/8 and the most common CVE being CVE-2017-0144. These are well correlated as the NSA created CVE-2017-0144/Eternal Blue as a windows exploit targeting Microsoft's implementation of the Server Message Block (SMB) Protocol [27].

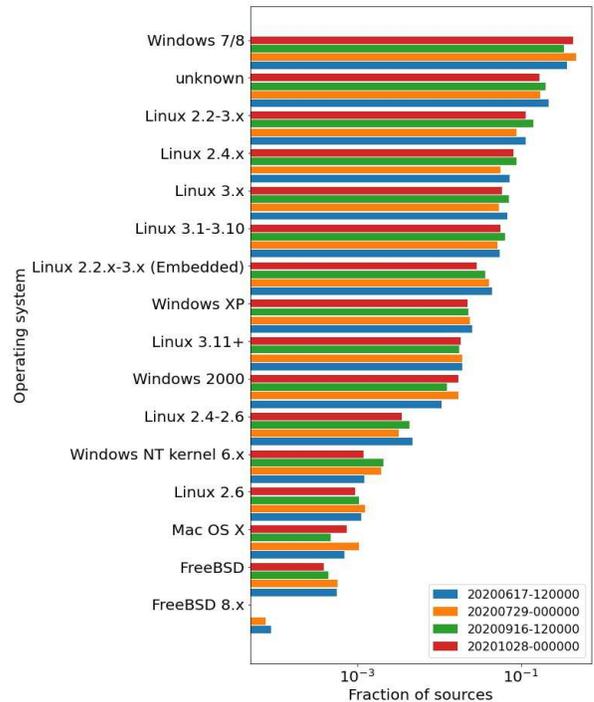

Figure 7. Distribution of the 16 most common operating systems observed by CAIDA and GreyNoise.

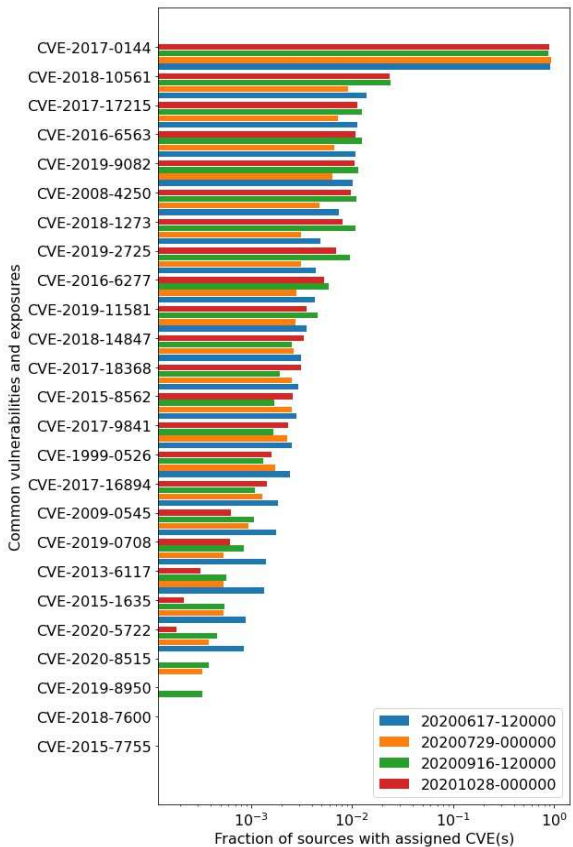

Figure 8. Distribution of the 25 most common CVEs observed by CAIDA and GreyNoise. Data were normalized and arranged in descending order from top to bottom.

## V. CONCLUSIONS AND FUTURE WORK

In this paper, we analyzed a vast amount of raw network traffic from modern sensors by applying correlation to find the main characteristic of the network traffic observed by the CAIDA telescope and GreyNoise honeyfarm. Python GraphBLAS and PyD4M analysis frameworks were used to cross-correlate two datasets from the telescope and a honeyfarm. Hence statistical understanding was obtained, both in terms of dominant characteristics of illegitimate traffic and overall statistical distributions exhibited by the resulting cross-correlated data.

These analyses revealed that certain adversarial activities dominate the traffic – the typical adversary is an unknown actor sending between 64 and 127 packets with malicious intent over protocol port TCP/455 using the Windows 7/8 OS and attempting to utilize CVE-2017-0144/Eternal Blue As such, cyber defenders should prioritize protection from this adversarial behavior.

Zipf-Mandelbrot distributions model several different phenomena arising from the examined illegitimate traffic, including the packet distribution sent from source IPs as well as the distribution of protocol ports and ASNs among source IPs. A result of these models is that few source IPs send large numbers of packets and that illegitimate traffic highly favors uncommon protocol ports and ASNs.

Due to the previously mentioned, network traffic is caused by small types of cyber activities. Knowing these characteristics, as the typical record, enables focus on the main threats based on statistical occurrences.

Future work will focus on the application of topic modeling techniques – such as sparse nonnegative matrix factorization and truncated singular value decomposition – to the incidence matrices associated with each relevant variable. Such analysis may reveal further information about the dimensionality of the data and the clustering of source IP-variable value pairs.

## VI. ACKNOWLEDGMENTS

The authors wish to acknowledge the following individuals for their contributions and support: Sean Atkins, Bob Bond, Ronisha Carter, K Claffy, Cary Conrad, Tucker Hamilton, Salim Hariri, Bill Hayes, Jeff Gottschalk, Raul Harnasch, Chris Hill, Mike Kanaan, Tim Kraska, Charles Leiserson, Mimi McClure, Sharon Oneal, Christian Prothmann, John Radovan, Steve Rejto, Daniela Rus, Doug Stetson, Scott Weed, Marc Zissman, and the MIT SuperCloud team: Bill Arcand, Bill Bergeron, David Bestor, Chansup Byun, Vijay Gadepally, Michael Houle, Matthew Hubbell, Anna Klein, Joseph McDonald, Peter Michaleas, Lauren Milechin, Julie Mullen, Andrew Prout, Antonio Rosa, Sid Samsi, Albert Reuther, Matthew Weiss, Charles Yee.

This material is based upon work supported by the Assistant Secretary of Defense for Research and Engineering under Air Force Contract No. FA8702-15-D-0001, National Science Foundation CCF-1533644, and United States Air Force Research Laboratory and Air Force Artificial Intelligence Accelerator Cooperative Agreement Number FA8750-19-2-1000. Any opinions, findings, conclusions, or recommendations expressed in this material are those of the author(s) and do not necessarily reflect the views of the Assistant Secretary of Defense for Research and Engineering, the National Science Foundation, or the United States Air Force. The U.S. Government is authorized to reproduce and distribute reprints for Government purposes, notwithstanding any copyright notation herein.

This work is partly supported by the National Science Foundation (NSF) research projects NSF-1624668, (NSF) DUE-1303362 (Scholarship-for-Service, and Department of Energy/National Nuclear Security Administration under Award Number(s) DE-NA0003946. Acknowledgments to CONACyT's Ph.D. fellowship program grant 786773.

In collaboration with the Information Security Research and Education (INSuRE) collaborative, a network of National Centers of Academic Excellence in Cyber Defense Research (CAE-R) universities that cooperate to engage students in solving applied cybersecurity research problems. (Arminda Estrada and Ivan Kawaminami are co-first authors).

# Enriquecimiento a gran escala y caracterización cibernética estadística del tráfico de red

Ivan Kawaminami[1], Arminda Estrada[1], Youssef Elsakkary[1], Hayden Jananthan[2], Aydın Buluç[3], Tim Davis[4], Daniel Grant[5], Michael Jones[2], Chad Meiners[2], Sandeep Pisharody[2], Jeremy Kepner[2]

[1]University of Arizona, [2]MIT, [3]LBNL, [4]Texas A&M, [5]GreyNoise,

*Abstract—* **Los sensores de red modernos producen enormes cantidades de datos sin procesar que están más allá de la capacidad del análisis humano. Una correlación cruzada de sensores de red se convierte en un desafío al enriquecer cada evento de red con metadatos adicionales. Estos grandes volúmenes de datos de red enriquecidos presentan una oportunidad para caracterizar estadísticamente el tráfico de red y responder a la pregunta: "¿Cuáles son las principales características cibernéticas de mis datos de red?" Los esquemas de análisis de Python GraphBLAS y D4M permiten realizar análisis estadísticos anónimos, rápidos y eficientes en conjuntos grandes de datos de red. Este enfoque se prueba utilizando miles de millones de muestras de datos de red anónimos del observatorio de Internet más grande (Telescopio CAIDA) y decenas de millones de registros anónimos del fondo comercial con la mayor capacidad de enriquecimiento (GreyNoise). El análisis confirma que la mayoría de las variables enriquecidas siguen las distribuciones de cola pesada y que una gran fracción del tráfico de red se debe a una pequeña cantidad de actividades cibernéticas. Esta información puede simplificar la tarea de los analistas cibernéticos al permitir la priorización de las actividades cibernéticas en función de la prevalencia estadística.**

*Keywords*—**Ciberseguridad, Computación de alto rendimiento, Big Data, Escaneo de redes, Análisis dimensional, Modelado de Internet, Captura de paquetes, Gráficos de transmisión.**

## I. Introducción

"¿Cuáles son las principales características cibernéticas de mis datos de red?" Es una pregunta crítica para cualquiera que proteja una red. Comprender las características de los datos de red es una herramienta muy útil para priorizar los recursos en actividades que lo requieran en actividades cibernéticas. La caracterización cibernética a menudo requiere análisis y correlación de un volumen significativo de tráfico de red a menudo contiene miles de millones de registros (anonimizados).

El uso de sensores de Internet permite recopilar información de datos sin procesar que se utilizan para caracterizar el tráfico de red para ayudar a las organizaciones a proteger los activos cibernéticos.[1]. Este trabajo exhibe el uso de conjuntos de datos de el Telescopio del Centro para el Análisis de Datos de Internet Aplicados (CAIDA) y el sistema de señuelo tipo "honeyfarm" GreyNoise que se correlacionan de manera cruzada utilizando GraphBLAS (GraphBLAS.org) y la librería Python "Dynamic Distributed Dimensional Data Model" (D4M). [2].

El Telescopio CAIDA monitorea el tráfico que entra y sale de un espacio del Internet oscuro que es una red /8 enrutada globalmente que transporta un tráfico legítimo mínimo debido a las pocas direcciones asignadas en el prefijo de Internet [3]. Los sistemas de señuelo "honeyfarm" son un mecanismo para observar e interactuar con el tráfico de Internet, a diferencia del telescopio que solo puede observar el tráfico [4]. El sistema de señuelo GreyNoise consta de cientos de servidores que escanean Internet recopilando pasivamente paquetes IP. Los servidores GreyNoise tienen la capacidad adicional de conversar con los remitentes de los paquetes para identificar su comportamiento, los métodos y las intenciones [5,6]. Una diferencia significativa entre estos dos conjuntos de datos es que CAIDA recopila exponencialmente más datos durante el mismo período que GreyNoise: aproximadamente treinta minutos de recopilación de datos de CAIDA equivalen a treinta días de recopilación de datos de GreyNoise. La tabla I muestra los datos recopilados utilizados para esta investigación. Los cuales tienen dos horas de inicio, GreyNoise y CAIDA. A lo largo de este documento, la hora de inicio de CAIDA se utiliza para especificar las fechas de recolección de datos, como se ve en las Tablas II-VI.

Tabla I. Hora de inicio de la recopilación de datos, duración de la recopilación y número de remitentes únicos de los conjuntos de datos de GreyNoise y CAIDA. Los datos de GreyNoise se recopilaron durante un mes, mientras que los datos de CAIDA se recopilaron aproximadamente cada seis semanas los miércoles, ya sea al mediodía o a la medianoche. Las muestras de paquetes constantes y de tiempo variable simplifican el análisis estadístico de las distribuciones de cola pesada que se encuentran comúnmente en el tráfico de red. [7].

| Inicio GreyNoise | Duración GreyNoise | Remitentes GreyNoise | Inicio CAIDA | Duración CAIDA | Paquetes CAIDA | Remitentes CAIDA |
|---|---|---|---|---|---|---|
| 2020-06-01 | 30 days | 1,111,458 | 20200617-120000 | 1594 sec | $2^{30}$ | 670,304 |
| 2020-07-01 | 31 days | 1,438,698 | 20200729-000000 | 1312 sec | $2^{30}$ | 541,300 |
| 2020-09-01 | 30 days | 1,245,194 | 20200916-120000 | 997 sec | $2^{30}$ | 723,991 |
| 2020-10-01 | 31 days | 1,997,782 | 20201028-000000 | 1068 sec | $2^{30}$ | 796,327 |

Los conjuntos de datos contienen tráfico de red que se encuentra al recopilar paquetes de datos de miles de direcciones de Protocolo de Internet (IP) que escanean Internet todos los días. El trabajo anterior [7] examinó la correlación temporal entre las fuentes CAIDA vistas por GreyNoise durante el mismo mes y encontró que entre los remitentes de CAIDA que han enviado paquetes fuente $d < N_V^{1/2}$, la fracción encontrada en los remitentes de GreyNoise se describe mediante una modificación escalada de la distribución de Cauchy.

$$p(d,t) \propto \frac{log_2(d)}{log_2(N_V^{1/2})} \frac{\beta}{\beta + |t - t_0|^\alpha}$$

donde $t$ es el tiempo de medición de GreyNoise, $t_0$ es el tiempo de medición de CAIDA, $N_V$ es el número de paquetes válidos consecutivos entre tiempos de medición $\alpha$ y $\beta$ son hiper parámetros independientes de $d$ y $t$. En general, el 70% de los registros más frecuentes ($d > N_V^{1/2}$) en CAIDA son observados en GreyNoise durante la misma ventana de tiempo.

Sobre la base del trabajo anterior, este documento explora en detalle las características cibernéticas de los registros correlacionados que se ven en ambos conjuntos de datos. Comprender las características del tráfico de red podría ayudar a defenderse en contra de ataques cibernéticos. Los datos del espacio oscuro del telescopio CAIDA son un rico punto de partida para esta caracterización, ya que representa casi en su totalidad el tráfico adversario [8-10]. Trabajos anteriores se realizaron en colaboración con diferentes centros de recopilación de datos, como el proyecto Widely Integrated Distributed Environment (WIDE) y CAIDA, con publicaciones relacionadas con el estado global del tráfico de Internet y prueba de prototipos para detener gusanos de Internet [11,12].

En trabajos relacionados, la caracterización de redes puede aprovechar aprendizaje automático (Machine Learning) para detección de anomalías, detección de intrusiones, y análisis de comportamiento [13]. Algunos clasificadores de aprendizaje automático dan como resultado una predicción de actividad maliciosa con un 96 % de precisión para la detección de intrusiones, incluso en vulnerabilidades explotadas en ataques de día cero en el tráfico de Internet [14]. Los datos de tráfico de Internet se pueden clasificar en tipos, según transmisión de audio, navegación, P2P, transferencia de archivos, transmisión de video y VOIP, además según las características de clasificación representativas, como la duración de flujo de datos, los intervalos de tiempo entre paquetes, las estadísticas de paquetes, la cantidad de paquetes, paquetes entrantes, paquetes salientes, etc. Cada categoría tiene ventajas y desventajas en los algoritmos implementados, lo que dificulta clasificar dos ataques en la misma categoría [15-19]. Los enfoques incluyen redes neuronales convolucionales y redes bidireccionales de memoria a corto plazo para capturar características en el contenido del paquete. Lo cual dio como resultado una precisión en la multi clasificación del 92 % para los tipos de tráfico cifrado [20].

CAIDA se ha utilizado comúnmente para comprender el tráfico de Internet, en [21-24] se aborda el problema de ubicar el Número de Sistema Autónomos (ASN) exactos, que controlan el enrutamiento de Internet. En este caso, se discrimina el tráfico analizando todos los niveles y caminos posibles provenientes de los AS al combinar información obtenida a través del uso de CAIDA; además de analizar la gran escala de datos de tráfico mediante el aprendizaje automático, es posible crear sistemas para la detección de intrusos. Además, en [25] el tráfico obtenido de CAIDA fue utilizado para observar el comportamiento de virus y gusanos a lo largo del tiempo aplicando enfoques mencionados anteriormente.

El trabajo previo se ha centrado en la clasificación específica de datos de tráfico de red, pero debido a limitaciones computacionales, la caracterización estadística general a gran escala ha sido un desafío. Este artículo demuestra la factibilidad de tal caracterización incorporando miles de millones de paquetes y 100,000 de fuentes (ver Tabla I). Además, el Protocolo de Internet (IP) con direcciones anónimas se utiliza para demostrar que tales caracterizaciones se pueden hacer con consideraciones de privacidad. Trabajar con direcciones IP anónimas se convierte en una herramienta útil en el intercambio de dichos datos. Anonimizar las direcciones IP permite compartir los datos manteniendo la privacidad de la información del usuario. Los marcos de análisis Python GraphBLAS y PyD4M son herramientas clave para realizar de manera eficiente y eficaz este análisis en grandes cantidades de datos anonimizados.

Esta investigación aborda la correlación cruzada de los conjuntos de datos CAIDA/GreyNoise en el siguiente orden: la Sección II presenta un análisis dimensional creado para determinar las variables de interés. La Sección III contiene un registro ejemplar del conjunto de datos analizado por correlación cruzada para las variables exploradas. La Sección IV muestra las distribuciones de probabilidad de variables seleccionadas y las modela con una distribución Zipf-Mandelbrot de 2 parámetros. Finalmente, la Sección V presenta las conclusiones del análisis y el trabajo futuro.

## II. ANÁLISIS DIMENSIONAL

El análisis dimensional se utiliza para observar las relaciones entre cantidades en función de sus dimensiones en sistemas complejos. Provee una manera de comprender el contenido de la información de las bases de datos y encontrar inconsistencias, patrones de datos y formatos. Las relaciones no cambian cuando se alteran las unidades de medida [26]. Este concepto se ha extendido a "big data" en forma de un análisis dimensional de datos que observa el tamaño (o dimensión) de diferentes valores en los datos [27]. En este trabajo, el análisis dimensional de datos de CAIDA y GreyNoise ayuda a identificar aquellas variables que requieren de una exploración detallada. Dada la cantidad y el tipo de datos que se manejan, se utiliza la biblioteca de Python D4M, que usa los conceptos de álgebra lineal y procesamiento de señales para bases de datos a través de matrices asociativas. Así proporciona un esquema capaz de representar la mayoría de los datos y con un acceso simple a través de una API.

Los datos de CAIDA y GreyNoise se construyen al crear matrices de tráfico hiper disperso en GraphBLAS para cada uno de los $2^{30}$ paquetes muestreados en un conjunto numerado de CAIDA en la Tabla I. El número de paquetes recibidos agrupado logarítmicamente, $\log_2(d)$, de cada registro CAIDA es calculado y las direcciones IP de remitentes se anonimiza utilizando el protocolo CryptoPAN. Asimismo, las IP de remitentes de GreyNoise se anonimizan de manera similar, lo que permite que los dos conjuntos de datos puedan estar correlacionados.

Las variables combinadas CAIDA y GreyNoise son asociadas con cada IP de origen anónimo son entonces separadas en tres tablas: una que contiene variables que no requieren más análisis y dos tablas con variables que requieren más análisis. Las últimas dos tablas corresponden a cuántos valores únicos están asociados con esas variables, ya sea alto (~10,000 o más, Tabla IV) o bajo (~10, Tabla V). Las tres tablas tienen el mismo formato general con encabezados de columna:

**Fecha:** la fecha y hora durante la cual CAIDA recopiló datos, aproximadamente cada seis semanas los miércoles, ya sea al mediodía (120000) o a la medianoche (000000).

**Variable:** la variable de interés para el análisis, recopilada por CAIDA y GreyNoise.

**Filas:** El número de filas asociadas con esta variable.

**Columnas:** el número de valores únicos para variables específicas.

**Vdc:** el número de valores distintos de cero.

**Maxval:** el valor que aparece con mayor frecuencia.

**Maxcont:** el número de veces que aparece el valor.

*Maxfrac*: la fracción del número de veces que aparece el valor máximo sobre el número total de valores de cada variable (es decir, Maxfrac = Maxcont/Columnas). Maxfrac se expresa en cursiva para indicar que se puede determinar directamente a partir de Maxcont y Columnas.

A partir de los datos proporcionados por CAIDA y GreyNoise, las variables elegidas para su análisis incluyen:

**ASN:** Número de Sistema Autónomo es el número que identifica a cada sistema autónomo. Es una colección de prefijos de enrutamiento de protocolo de Internet conectados controlados por operadores de red.

**Clasificación:** el tráfico de Internet se comunica a través de paquetes; GreyNoise clasifica esos paquetes como desconocidos, benignos y maliciosos. Un dispositivo puede clasificarse como "malicioso" si se marca como que tiene intenciones maliciosas, "benigno" si no tiene intenciones maliciosas y "desconocido" si se desconoce el propósito del paquete transmitido.

**CVE:** Enumeración Común de Vulnerabilidades (CVE) enumera vulnerabilidades conocidas públicamente en dispositivos con información relacionada. Los registros clasificados como "benignos" o "desconocidos" no tienen un valor CVE, mientras que cada registro "malicioso" puede tener cero o más CVE asociados.

**Última Marca de Tiempo (UMT):** la marca de tiempo cuando GreyNoise observó el paquete por última vez. El tiempo proporciona información sobre todas las interacciones y el comportamiento de todos los paquetes analizados.

**OS:** El sistema operativo (OS) es el software que administra los recursos de hardware y las funciones esenciales de otras computadoras. El sistema operativo proviene de la IP del remitente.

**Puerto:** este número identifica el estándar utilizado para comunicarse con un puerto de computadora.

**Paquetes de origen**: cantidad de paquetes enviados desde la IP de origen, agrupados en contenedores de tamaños 2n.

**Paquete de origen No Grey:** la cantidad de paquetes enviados desde direcciones IP de origen de CAIDA únicamente, agrupados en contenedores de tamaños 2n.

**Actor:** participante que envía paquetes a través de la red.

**CAIDA Grey Meta/ CAIDA No Grey:** los valores totales en la matriz para cada archivo, donde caidaGrayMeta son los datos correlacionados entre GreyNoise y CAIDA y caidaNoGrey son datos obtenidos únicamente de CAIDA.

**Spoofable:** esta variable binaria identifica si la IP de origen no pudo completar una conexión TCP completa o no.

*A. Variables de origen*

La Tabla II contiene información más detallada sobre la cantidad de datos obtenidos. La columna Fecha muestra cuatro fechas diferentes que comienzan con el año, el mes y el día del mes. Si una fecha termina en "120000", los datos fueron recopilados por CAIDA al mediodía; sin embargo, si la fecha termina en "000000", los datos se recopilaron a medianoche.

Tabla II. Tabla de origen de los datos recopilados en cuatro fechas diferentes. La fecha se divide en año, mes y día del mes. Las variables caidaGreyMeta corresponden a las IP de origen que aparecen en los conjuntos de datos de GreyNoise y CAIDA, mientras que caidaNoGrey corresponde a las IP de origen que aparecen solo en el conjunto de datos de CAIDA.

| Fecha | Variable | Filas | Columnas | Vdc | max count | *maxfrac* |
|---|---|---|---|---|---|---|
| 20200617-120000 | caidaNoGrey | 545,980 | 1 | 545,980 | 545,980 | 1 |
| 20200729-000000 | caidaNoGrey | 394,875 | 1 | 394,875 | 394,875 | 1 |
| 20200916-120000 | caidaNoGrey | 589,267 | 1 | 589,267 | 589,267 | 1 |
| 20201028-000000 | caidaNoGrey | 603,273 | 1 | 603,273 | 603,273 | 1 |
| 20200617-120000 | caidaGreyMeta | 124,300 | 9 | 942,112 | 124,300 | 0.131 |
| 20200729-000000 | caidaGreyMeta | 146,402 | 9 | 1,118,893 | 146,402 | 0.130 |
| 20200916-120000 | caidaGreyMeta | 134,699 | 9 | 1,034,353 | 134,699 | 0.130 |
| 20201028-000000 | caidaGreyMeta | 193,029 | 9 | 1,470,649 | 193,029 | 0.131 |

**caidaNoGrey/caidaGreyMeta:** La Tabla II contiene la información de los metadatos de los registros con correlación cruzada de CAIDA-GreyNoise y los metadatos para los registros de CAIDA sin metadatos de GreyNoise asociados. Para caidaGreyMeta, hay nueve valores distintos que representan las variables que se analizan: actor, spoofable, ASN, última marca de tiempo, puerto, OS, srcPacket, clasificación y CVE. Por otro lado, caidaNoGrey muestra la única variable srcPacketNoGrey.

*B. Variables Irrelevantes*

La Tabla III muestra las variables con un número limitado de valores distintos que es poco probable que brinden información adicional.

**Spoofable:** La Tabla III muestra una correlación de muchos a uno entre los paquetes spoofable y el valor booleano "1". Entre el 30 % y el 40 % de los paquetes son falsificables, aunque estar marcados como falsificados no implica necesariamente que el tráfico haya sido falsificado, solo que no se pudo establecer lo contrario. No parece haber ninguna correlación importante entre ser marcado como suplantable y las variables restantes, por lo que la variable se considera irrelevante para un análisis posterior.

**Actor:** La Tabla III muestra que se encuentran entre 124.300 y 193.029 registros para los actores con una correlación de muchos a uno; en otras palabras, para cada paquete, hay un solo actor (aunque este actor puede ser "desconocido"). Hay entre 22 y 30 actores únicos, siendo el actor más común "desconocido", el cual aparece aproximadamente el 99% del tiempo. Debido a que muy pocos registros tienen un actor "conocido", esta variable no es relevante para nuestro estudio de los patrones en los datos.

Tabla III. Información concentrada y relacionada con variables irrelevantes, donde la Fecha específica cuando recolectaron los datos. Las variables incluyen actor y spoofable, siendo el valor máximo de actor "desconocido" para cualquier fecha, mientras que spoofable contiene entre 36 000 y 63 000 instancias.

| Fecha | variable | Filas | Columnas | Vdc | maxval | max count | maxfrac |
|---|---|---|---|---|---|---|---|
| 20200617-120000 | spoofable | 36,317 | 1 | 36,317 | 1 | 36,317 | 1 |
| 20200729-000000 | spoofable | 43,838 | 1 | 43,838 | 1 | 43,838 | 1 |
| 20200916-120000 | spoofable | 56,189 | 1 | 56,189 | 1 | 56,189 | 1 |
| 20201028-000000 | spoofable | 62,172 | 1 | 62,172 | 1 | 62,172 | 1 |
| 20200617-120000 | actor | 124,300 | 22 | 124,300 | desconocido | 123,843 | 0.996 |
| 20200729-000000 | actor | 146,402 | 26 | 146,402 | desconocido | 145,812 | 0.996 |
| 20200916-120000 | actor | 134,699 | 30 | 134,699 | desconocido | 133,036 | 0.987 |
| 20201028-000000 | actor | 193,029 | 29 | 193,029 | desconocido | 192,091 | 0.995 |

### C. Variables Relevantes

La Tabla IV y la Tabla V están separadas por variables con similitud estadística, el tamaño del número de columnas y, en particular, el número de valores únicos en los registros. La Tabla IV contiene las variables con muchos valores únicos en los miles, mientras que la Tabla V contiene varios valores únicos de solo dos dígitos. Las variables en ambas tablas serán analizadas usando distribuciones de probabilidad e histogramas.

Tabla IV. Variables relacionadas a muchos (~10 000 y más) valores únicos, cuyo número está representado por Columnas. Las variables incluyen ASN, última marca de tiempo y puerto.

| Fecha | Variable | Filas | Columnas | Vdc | maxval | maxcont | maxfrac |
|---|---|---|---|---|---|---|---|
| 20200617-120000 | puerto | 124,300 | 4,021 | 203,081 | TCP/445 | 44,607 | 0.219 |
| 20200729-000000 | puerto | 146,402 | 4,353 | 219,650 | TCP/445 | 66,155 | 0.301 |
| 20200916-120000 | puerto | 134,699 | 6,075 | 210,315 | TCP/23 | 46,874 | 0.223 |
| 20201028-000000 | puerto | 193,029 | 5,541 | 304,387 | TCP/445 | 78,971 | 0.259 |
| 20200617-120000 | ASN | 124,300 | 10,304 | 124,300 | AS4134 | 6,096 | 0.049 |
| 20200729-000000 | ASN | 146,402 | 9,825 | 146,402 | AS3462 | 8,650 | 0.059 |
| 20200916-120000 | ASN | 134,699 | 10,059 | 134,699 | AS17488 | 130,24 | 0.097 |
| 20201028-000000 | ASN | 193,029 | 10,764 | 193,029 | AS4837 | 118,30 | 0.061 |
| 20200617-120000 | UMT | 124,300 | 115,900 | 124,300 | 6/23/20 23:20 | 18 | 0.000143 |
| 20200729-000000 | UMT | 146,402 | 130,474 | 146,402 | 8/1/20 0:09 | 16 | 0.000109 |
| 20200916-120000 | UMT | 134,699 | 127,504 | 134,699 | 10/1/20 0:11 | 10 | 0.000074 |
| 20201028-000000 | UMT | 193,029 | 157,549 | 193,029 | 10/31/20 23:12 | 40 | 0.000207 |

**Puerto:** algunas veces, una IP de origen usaba muchos puertos para la transmisión de un solo paquete. El número de puertos por registro se limitó a cinco para preservar la viabilidad de trabajar con los datos. Los puertos únicos estaban entre 40.000 y 61.000. TCP/445 (SMB) fue el más frecuente en tres de las fechas de recolección y TCP/23 en 20200916-120000. El puerto de protocolo más popular aparece entre el 21 % y el 30 % de las veces.

**ASN:** Cada registro tiene un ASN asociado. Hay miles de ASN distintos y hay un valor máximo diferente para cada una de las fechas. Las veces que se contabilizan esos valores máximos oscilan entre 6,096 y 13,024 veces, lo que corresponde a una fracción entre el 4.9% y el 6.1% del total de ASN registrados.

**Última Marca de Tiempo (UTM):** Para cada registro dado, hay exactamente una última marca de tiempo vista. Hay entre 115 900 y 157 549 marcas de tiempo únicas vistas por última vez. Los valores máximos, como era de esperar, están relacionados con la fecha y la hora en que se recopilaron los datos. El recuento del rango de marcas de tiempo máximas está entre 10 y 40, o menos del 0.02 % del número total de marcas de tiempo registradas.

La Tabla V contiene variables que serán analizadas más adelante. Las similitudes entre estas variables se definen por el número de valores únicos (Columnas) que es bastante bajo, entre tres y 36 valores únicos cada uno.

Tabla V. Variables que exhiben relativamente pocos valores únicos (~10), incluida la clasificación, CVE, OS, srcPacket y srcPacketnoGrey.

| date | variable | Filas | Columnas | Vdc | maxval | max cont | max frac |
|---|---|---|---|---|---|---|---|
| 20200617-120000 | CVE | 35,695 | 36 | 37,150 | 2017-0144 | 33,660 | 0.906 |
| 20200729-000000 | CVE | 50,241 | 33 | 51,546 | 2017-0144 | 48,265 | 0.936 |
| 20200916-120000 | CVE | 35,271 | 32 | 36,500 | 2017-0144 | 32,205 | 0.882 |
| 20201028-000000 | CVE | 57,274 | 34 | 59,252 | 2017-0144 | 52,880 | 0.892 |
| 20200617-120000 | clasificación | 124,300 | 3 | 124,300 | malicioso | 72,952 | 0.587 |
| 20200729-000000 | clasificación | 146,402 | 3 | 146,402 | malicioso | 86,058 | 0.588 |
| 20200916-120000 | clasificación | 134,699 | 3 | 134,699 | malicioso | 86,141 | 0.639 |
| 20201028-000000 | clasificación | 193,029 | 3 | 193,029 | malicioso | 109,849 | 0.569 |
| 20200617-120000 | OS | 124,300 | 17 | 124,300 | windows 7/8 | 45,107 | 0.363 |
| 20200729-000000 | OS | 146,402 | 20 | 146,402 | windows 7/8 | 68,856 | 0.470 |
| 20200916-120000 | OS | 134,699 | 31 | 134,699 | windows 7/8 | 45,502 | 0.338 |
| 20201028-000000 | OS | 193,029 | 20 | 193,029 | windows 7/8 | 83,013 | 0.430 |
| 20200617-120000 | caida srcpacket | 545,980 | 23 | 545,980 | 32 | 104,466 | 0.191 |
| 20200729-000000 | caida srcpcaket | 394,875 | 24 | 394,875 | 32 | 67,020 | 0.169 |
| 20200916-120000 | caida srcpacket | 589,267 | 22 | 589,267 | 1 | 113,371 | 0.1925 |
| 20201028-000000 | caida srcpacket | 603,273 | 24 | 603,273 | 1 | 105,623 | 0.175 |
| 20200617-120000 | src packets | 124,300 | 25 | 124,300 | 128 | 20,570 | 0.165 |
| 20200729-000000 | src packets | 146,402 | 24 | 146,402 | 64 | 19,750 | 0.135 |
| 20200916-120000 | src packets | 134,699 | 26 | 134,699 | 64 | 21,907 | 0.166 |
| 20201028-000000 | src packets | 193,029 | 25 | 193,029 | 64 | 31,783 | 0.165 |

**CVE:** entre 35.271 y 57.274 registros tienen un CVE, equivalente al 26-30% del tráfico malicioso representado, lo cual es menos de la mitad de los registros. Muchos CVE pueden aparecer en un solo registro. El CVE más popular en todas las fechas es CVE-2017-0144. Se trata de una conocida vulnerabilidad llamada Eternal Blue relacionada con el famoso ataque WannaCry que se convirtió en un problema mundial en 2017 [28]. Aparece entre el 88% y el 93% de las veces que se registró un CVE.

**Clasificación:** hay tres valores únicos: "malicioso", "benigno" y "desconocido". Entre el 59.6 % y el 63.95 % de los registros (lo que representa del 72,952 al 109,849 de los registros) son clasificados como "maliciosos".

**OS:** El número de registros con sistema operativo reportado se encuentra entre 124,300 y 193,029. Cada registro tiene valor de sistema operativo registrado, aunque esto incluye "desconocido", lo que indica que GreyNoise no pudo determinar el sistema operativo de origen. Existe una relación de muchos a uno entre las direcciones IP de origen y el sistema operativo, con varios valores únicos que van de 17 a 31 (incluido "desconocido"); como se observa en la Tabla IV. El sistema operativo más común es Windows 7/8, con un recuento máximo que va de 45,107 a 83,013, lo que representa una fracción entre el 33 % y el 47 % de las instancias registradas.

**Packet de origen No Grey (caida srcpacket):** los registros con paquetes de origen reportado oscilan entre 39,487 y 60,327.

La Tabla IV muestra una correlación de muchos a uno entre las direcciones IP de origen y el número (agrupado) de paquetes de origen encontrados en los registros de CAIDA. El tamaño de contenedor de paquetes de origen más común es 1 para dos fechas y 32-63 para las otras dos fechas. La cantidad máxima de veces que aparece un rango de tamaño de paquete de origen específico es de 67,020 a 113,371 veces o del 16.98 % al 19.24 % de las veces.

**Packet de origen (src packet):** los registros con paquetes de origen están entre 124,300 y 193,029. Las fuentes únicas agrupadas en tamaños 2n oscilan entre 23 y 26. La Tabla V muestra una correlación de muchos a uno entre las direcciones IP de origen y los contenedores de paquetes de origen; para cada registro, se encuentra exactamente un rango de paquetes de origen. El tamaño de paquete de origen más común está entre 64 y 127 kilobytes y, en una ocasión, entre 128 y 256 kilobytes. Entre el 13% al 16.5% de los paquetes pertenecen a este rango de tamaño o entre 19,750 a 31,783 registros.

### III. REGISTRO EJEMPLAR

Con base en el análisis de datos dimensionales, es posible dar una respuesta inicial a la pregunta "¿Cuáles son las características cibernéticas primarias de los datos de mi red?" en la forma de un registro ejemplar basado en los valores más comunes de los datos.

Tabla VI. Registro típico refleja los valores más frecuentes de cada variable.

| variable | maxval | Maxcont | *maxfrac* | date |
|---|---|---|---|---|
| actor | desconocido | 123843 | 0.996 | 20200617-120000 |
| | | 145812 | 0.996 | 20200729-000000 |
| | | 133036 | 0.988 | 20200916-120000 |
| | | 192091 | 0.995 | 20201028-000000 |
| CVE | CVE-2017-0144 | 33660 | 0.906 | 20200617-120000 |
| | | 48265 | 0.936 | 20200729-000000 |
| | | 32205 | 0.882 | 20200916-120000 |
| | | 52880 | 0.892 | 20201028-000000 |
| clasificación | malicioso | 72952 | 0.587 | 20200617-120000 |
| | | 86058 | 0.588 | 20200729-000000 |
| | | 86141 | 0.640 | 20200916-120000 |
| | | 109849 | 0.569 | 20201028-000000 |
| OS | Windows 7/8 | 45107 | 0.363 | 20200617-120000 |
| | | 68856 | 0.470 | 20200729-000000 |
| | | 45502 | 0.338 | 20200916-120000 |
| | | 83013 | 0.430 | 20201028-000000 |
| puerto | TCP/445 | 44607 | 0.220 | 20200617-120000 |
| | | 66155 | 0.301 | 20200729-000000 |
| | TCP/23 | 46874 | 0.223 | 20200916-120000 |
| | TCP/445 | 78971 | 0.259 | 20201028-000000 |
| Packet de origen noGrey | 32 | 104466 | 0.191 | 20200617-120000 |
| | | 67020 | 0.170 | 20200729-000000 |
| | 1 | 113371 | 0.192 | 20200916-120000 |
| | | 105623 | 0.175 | 20201028-000000 |
| Packet de origen | 128 | 20570 | 0.165 | 20200617-120000 |
| | 64 | 19750 | 0.135 | 20200729-000000 |
| | | 21907 | 0.163 | 20200916-120000 |
| | | 31783 | 0.165 | 20201028-000000 |
| ASN | AS4134 | 6096 | 0.049 | 20200617-120000 |
| | AS3462 | 8650 | 0.059 | 20200729-000000 |
| | AS17488 | 13024 | 0.097 | 20200916-120000 |
| | AS4837 | 11830 | 0.061 | 20201028-000000 |

La Tabla VI muestra según las variables, un registro típico basado en maxval y maxcount sería que el 99 % de las veces, un actor desconocido que ejecuta el sistema operativo Windows 7/8 envía entre 64 y 127 paquetes con intenciones maliciosas a través de TCP/455 con un ASN no específico y apunta a la vulnerabilidad CVE-2017-0144. Tal registro ejemplar proporciona operaciones de red con información altamente procesable sobre las actividades antagónicas más comunes en su red.

### IV. DISTRIBUCIÓN DE PROBABILIDAD

Mientras que un registro ejemplar proporciona una indicación de los más actividades comunes, las distribuciones de probabilidad arrojan luz sobre el comportamiento más amplio de la red. Además, la mayoría de las variables son consistentes con una distribución Zipf-Mandelbrot que permite la caracterización de los datos con tan solo 2 parámetros [29]. El modelo Zipf-Mandelbrot modificado no normalizado se denota por la siguiente ecuación:

$$\rho(d; \alpha, \delta) = \frac{1}{(d + \delta)^\alpha}$$

Donde d es el conteo del valor particular de la variable $\delta$ y $\alpha$ son parámetros encontrados usando los procedimientos de [7]. En las Figuras 1-4, se representa gráficamente el modelo modificado Zipf-Mandelbrot calculado, observando que el modelo $\rho(d;\alpha,\delta)$ está agrupado logarítmicamente antes de la representación gráfica. La Figura 1 muestra que la mayoría de las IP de origen de la red enviaron alrededor de $10^2$ paquetes, mientras que muy pocas IP de origen enviaron una gran cantidad de paquetes.

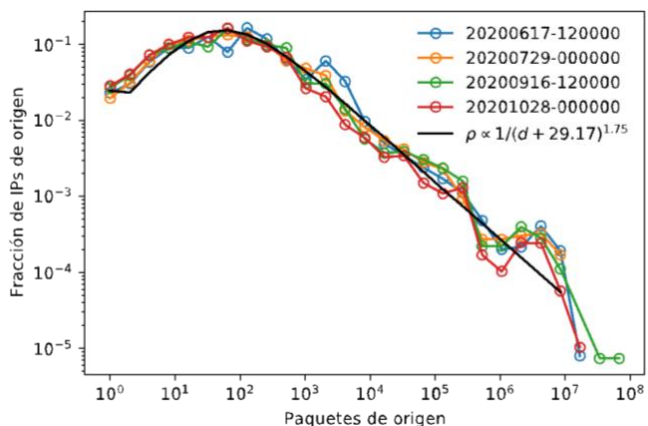

Figura 1. Distribución de probabilidad de la fracción de las IPs de origen de CAIDA (sin coincidencia con GreyNoise) respecto al número de paquetes de origen de la IP de origen. La mayoría de los paquetes de origen transmitieron entre $10^1$ y $10^2$ paquetes. El modelo Zipf-Mandelbrot modificado asociado se traza adicionalmente, con parámetros de modelo α=1.75 y δ=29.17.

La Figura 2 muestra un comportamiento similar en la variable de paquetes de origen noGrey, además de que CAIDA y GreyNoise están observando un ruido de fondo en Internet.

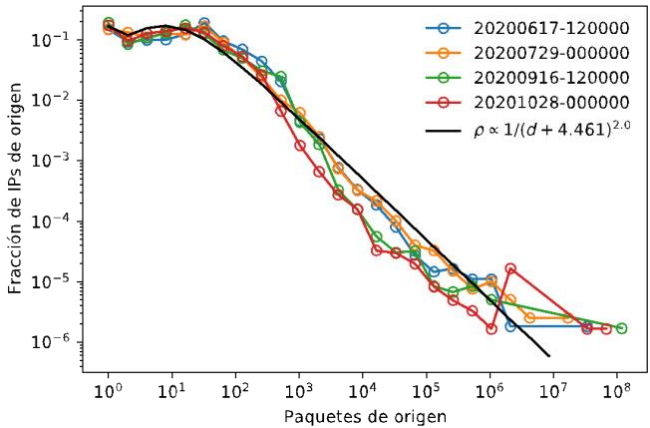

Figura 2. Distribución de probabilidad de la fracción de IP de origen de CAIDA (con coincidencias con GreyNoise) frente al número de paquetes según de la IP de origen. La mayoría de las IP de origen transmitieron menos de $10^2$ paquetes. Los datos son consistentes con la distribución Zipf-Mandelbrot con α=2.0 y δ=4.461. La distribución muestra una gran cantidad de direcciones IP de origen con una cantidad baja de paquetes de origen y un bajo número de direcciones IP de origen con un alto número de paquetes de origen.

El histograma de la Figura 3 muestra como los puertos de baja frecuencia representan una fracción mayor de los datos en comparación con los puertos de alta frecuencia, lo que sugiere que el tráfico ilegítimo favorece los puertos poco comunes en lugar de los puertos de alta frecuencia.

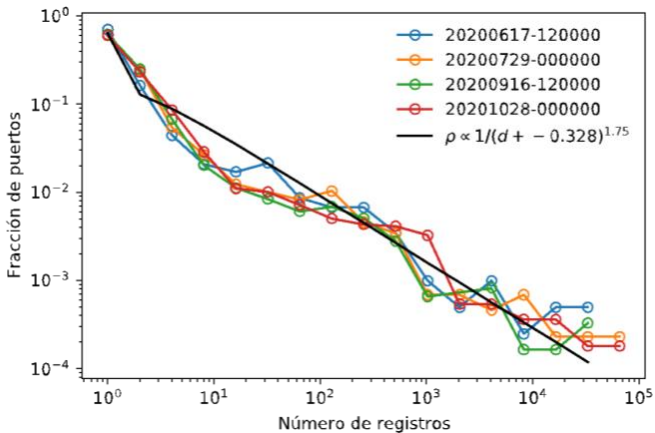

Figura 3. Distribución de probabilidad de la fracción de puertos frente al número de fuentes con ese puerto. Los datos son consistentes con la distribución Zipf-Mandelbrot con α=1.75 y δ=-0.328. Distribución de probabilidad del número de direcciones IP de origen frente a la fracción de puertos utilizados por ese número de fuentes. La distribución muestra un alto número de puertos utilizados por solo unas pocas fuentes y pocos puertos utilizados por muchas fuentes.

El mismo comportamiento se observa en la variable ASN, como se observa en la Figura 4, lo que sugiere que la variable ASN tiene valores poco comunes que están siendo favorecidos por el tráfico ilegítimo. Las figuras 2, 3 y 4 sugieren que estas variables son relativamente consistentes en el tiempo y consistente con el Zipf-Mandelbrot distribuciones. Estos datos permiten a un operador de red priorizar los registros en función de su impacto. Asimismo, la comparación de nuevos los datos con la distribución esperada se puede utilizar para detectar cambios en el comportamiento que puedan sugerir medidas adicionales.

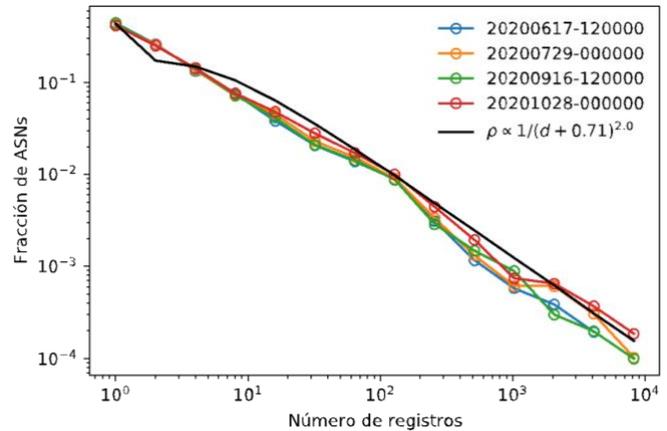

Figura 4. 4. Distribución de probabilidad de la fracción de ASN frente al número de registros con ese ASN. Los datos son consistentes con la distribución Zipf-Mandelbrot con α=2.0 y δ=0.72. La distribución muestra que muchos ASN fueron utilizados solo por unas pocas fuentes y pocos ASN fueron utilizados por muchos registros.

En la Figura 5, se muestra la correlación entre la hora de la marca de tiempo vista por última vez y la fracción de direcciones IP de origen vistas por última vez en dicha hora. La etiqueta 120000 explica que la recolección comenzó al mediodía, a diferencia de la etiqueta 000000 que indica que la recolección comenzó a la medianoche. Los datos recopilados a la misma hora muestran una estrecha correlación.

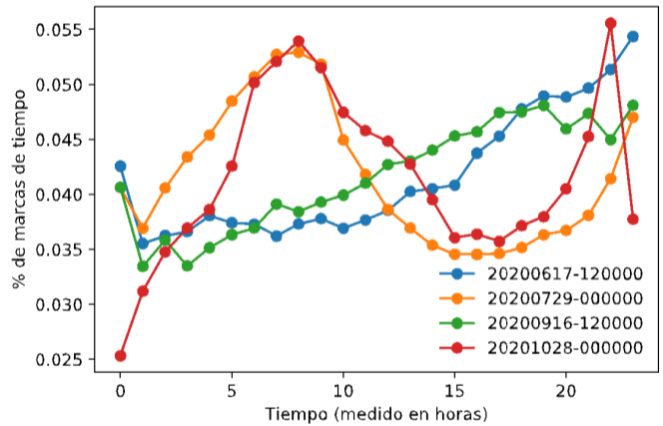

Figura 5. Distribución de probabilidad de la hora del día frente a la fracción de IP de origen cuya última marca de tiempo vista fue a esa hora del día. Para las fechas que terminan con 120000, los datos de CAIDA se recopilaron al mediodía; para fechas que terminan con 000000, la recolección fue a la medianoche. Las distribuciones de las últimas marcas de tiempo vistas son consistentes entre aquellos con los mismos tiempos de recolección de 120000 o 000000.

El histograma de clasificación en la Figura 6 muestra tres valores únicos, "benigno", "malicioso" y "desconocido", casi no hay paquetes benignos. Más bien, la mayoría son maliciosos y el resto son desconocidos. La frecuencia relativa es constante a lo largo del período de cuatro colecciones.

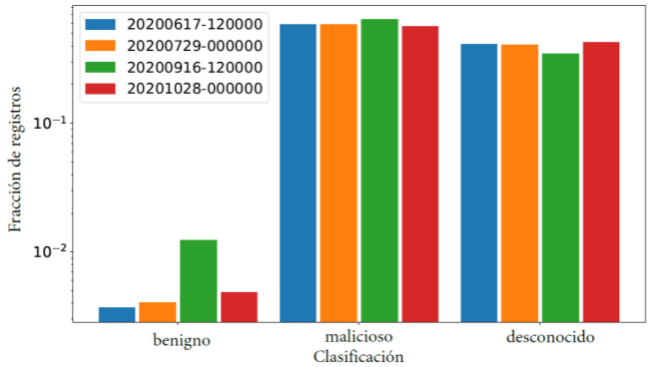

Figura 6. Frecuencia de tipo de paquete para CAIDA y GreyNoise. Existen tres tipos de clasificación, siendo la clasificación maliciosa la más frecuente en cada fecha.

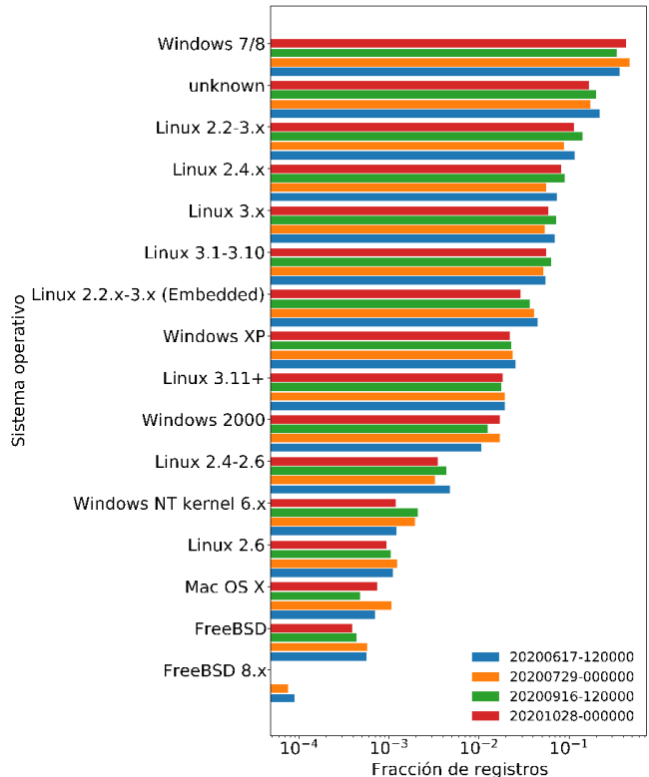

Figura 7. Distribución de los 16 sistemas operativos más comunes observados por CAIDA y GreyNoise.

La Figura 7 muestra una frecuencia que se distribuye uniformemente respecto a la fecha en que se recopilaron los datos. Si bien Windows 7/8 era el sistema operativo más común de la IP de origen, el segundo sistema operativo más común pertenece a sistemas operativos que no fueron reconocidos, seguido de cerca por varias versiones de Linux. La Figura 8 muestra una tendencia similar. Las figuras 7 y 8 corroboran el registro ejemplar mostrado en la Sección III, siendo el sistema operativo más común Windows 7/8 y el CVE más común CVE-2017-0144. Estos están bien correlacionados, ya que la Agencia de Seguridad Nacional (NSA) de los Estados Unidos creó la etiqueta CVE-2017-0144/Eternal Blue como una vulnerabilidad de Windows dirigido a la implementación de Microsoft del Protocolo Server Message Block (SMB) [27].

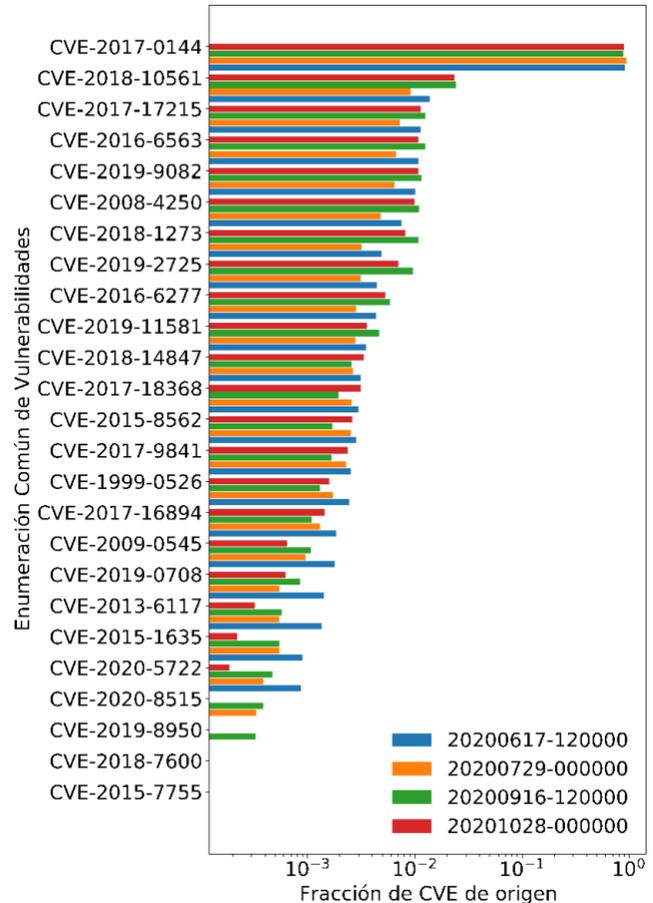

Figura 8. Frecuencias de los 25 CVE más comunes observados por CAIDA y GreyNoise. Los datos se normalizaron y organizaron en orden descendente desde la parte superior.

## V. Conclusiones y trabajo futuro

En este documento, analizamos una gran cantidad de tráfico de red sin procesar de sensores modernos mediante la aplicación de correlación para encontrar la característica principal del tráfico de red observado por el telescopio CAIDA y "honeyfarm" GreyNoise. Los esquemas de análisis en Python GraphBLAS y D4M se utilizaron para correlacionar dos conjuntos de datos del telescopio y sistema de señuelo. Por lo tanto, se obtuvo una comprensión estadística, tanto en términos de las características dominantes del tráfico ilegítimo como de las distribuciones estadísticas generales exhibidas por los datos de correlación cruzada resultantes.

Estos análisis revelaron que ciertas actividades adversarias dominan el tráfico: el adversario típico es un actor desconocido que envía entre 64 y 127 paquetes con intenciones maliciosas a través del puerto de protocolo TCP/455 utilizando el sistema operativo Windows 7/8 e intenta utilizar CVE-2017-0144/ Eternal Blue. Por lo tanto, los expertos en ciberseguridad deben priorizar la protección contra este comportamiento adversario.

Las distribuciones modificadas de Zipf-Mandelbrot modelan varios fenómenos diferentes que surgen del tráfico ilegítimo examinado, incluida la distribución de paquetes enviados desde

IP de origen, así como la distribución de puertos y ASN entre IP de origen. Un resultado de estos modelos es que pocas direcciones IP de origen envían una gran cantidad de paquetes; además, el tráfico ilegítimo favorece en gran medida los puertos poco comunes y los ASN.

Debido a lo mencionado anteriormente, el tráfico de red es causado por pequeños tipos de actividades cibernéticas. Conocer estas características, como el registro típico, permite enfocar las principales amenazas por ocurrencias estadísticas.

El trabajo futuro se centrará en la aplicación de técnicas de modelado de temas, como la factorización de matrices dispersas no negativas y la descomposición de valores singulares truncados, a las matrices de incidencia asociadas con cada variable relevante. Dicho análisis puede revelar más información sobre la dimensionalidad de los datos y la agrupación de pares de valores de variables IP de origen.

## VI. Agradecimientos